\documentclass[a4]{amsart}
\usepackage{amssymb}
\usepackage{amsfonts}
\usepackage{amstext}
\usepackage{graphicx}
\newcommand{\R}{{\mathbb{R}}}

\newcommand{\CD}{{\mathcal{D}}}
\newcommand{\CH}{{\mathcal{H}}}
\newcommand{\cm}{{\mathfrak{m}}}
\newcommand{\cg}{{\mathfrak{g}}}

\newcommand{\btens}{{\bar\otimes}}

\newcommand{\bd}{{\bar\extd}}
\newcommand{\bDelta}{{\bar\Delta}}

\newcommand{\extd}{{\rm d}}
\newcommand{\del}{{\partial}}

\newcommand{\bicross}{\triangleright\!\!\!\blacktriangleleft}
\newcommand{\lrbicross}{\blacktriangleright\!\!\!\triangleleft}
\newcommand{\lcross}{\rtimes}

\begin{document}

\title{Scaling Limit of the Noncommutative Black Hole}
\keywords{Quantum groups, noncommutative geometry, quantum gravity, black hole, redshift, time dilation, NASA Fermi satellite, quantum spacetime, bicrossproduct model}

%\subjclass{Primary 81R50, 16W50, 16S36}

\author{Shahn Majid}
\address{Queen Mary University of London\\
School of Mathematics, Mile End Rd, London E1 4NS, UK}
\thanks{The author was supported by a Leverhulme Senior Research Fellowship}
\email{s.majid@qmul.ac.uk}
\date{September 2010}

\begin{abstract} We show that the `quantum'  black hole wave operator in the $\kappa$-Minkowski or bicrossproduct model quantum spacetime introduced in \cite{Ma:bh} has a natural scaling limit $\lambda_p\to 0$ at the event horizon. Here $\lambda_p$ is the Planck time and the geometry at the event horizon in Planck length is maintained at the same time as the limit is taken, resulting in a classical theory with quantum gravity remnants.  Among the features is a frequency-dependent `skin' of some  $\omega\over\nu$ Planck lengths just inside the event horizon for $\omega>0$ and just outside for $\omega<0$, where $\nu$ is the frequency associated to the Schwarzschild radius. We use bessel and hypergeometric functions to analyse propagation through the event horizon and skin in both directions. The analysis confirms a finite redshift at the horizon for positive frequency modes in the exterior. \end{abstract}
\maketitle

\section{Introduction }

First applications of quantum groups and noncommutative geometry to model quantum gravity effects have been around for some 2 decades now\cite{Ma:pla}. The work \cite{Ma:qreg}  proposed how noncommutative spacetime might tame infinities in physics while in 1994 we introduced the particular bicrossproduct or $\kappa$-Minkowski spacetime \cite{MaRue} which has since attracted a fair amount of attention, mainly because it `almost commutative' and hence easier to work with. In this model the spatial generators $x_i$ mutually commute but the time variable $t$ does not. Rather,
  \begin{equation}\label{bicxt}  [x_i,t]=\imath\lambda_p x_i\end{equation}
 where $\lambda_p$ is a real parameter. If the noncommutative geometry arises as a model of quantum gravity corrections to classical geometry then presumably $\lambda_p\sim 10^{-44}s$ as Planck time.  The model has a bicrossproduct form of quantum Poincar\'e group and Section~2 provided a little more background to its construction.

In spite of its success, which includes the proposal of its  testing by  time of flight data being collected for gamma-ray bursts by FERMI-GLAST currently in orbit, the main criticism of the bicrossproduct model remains that it is rather special to flat spacetime, relying as it does on (quantum group) Fourier transform and plane waves\cite{AmeMa}. This does not sit too well as a quantum gravity effect as it effectively applies in some weak field limit. In \cite{Ma:bh} we have proposed an answer to this criticism; a general construction of a noncommutative spacetime  $C(M)\rtimes_\tau\R$ in which space is now any classical Riemannian 3-manifold $(M,\bar g)$ equipped with a conformal Killing vector field $\tau$ and a static metric of classical form 
\begin{equation}\label{static}\beta^{-1}\bd t\btens\bd t+\bar g\end{equation}
 where $\beta\in C(M)$ is a function on $M$. In our new theory the noncommutative differential geometry replaces the quantum Poincar\'e group as the guiding force, though the latter would be expected in the fullness of time to reemerge as the fiber of an associated affine quantum group frame bundle using the formalism of \cite{Ma:rie}.  In fact the noncommutative geometry  does not contain a full development of the noncommuative metric and noncommutative Riemannian geometry, although again this should emerge in the fullness of time. Rather, it takes a shortcut straight to the physically important noncommutative spacetime wave operator.

This theory constructs the noncommutative wave-operator on $C(M)\rtimes_\tau\R$ viewed as quantizing  any static spacetime. This includes the Schwarzschild black hole. However, computations are quite formidable and hence in practice we work with a simplified version in which we use the flat bicrossproduct model spacetime (\ref{bicxt}) but `bolt on' the nontrivial Riemannian 3-geometry by a process of minimal coupling. This allows us to treat the black hole using the flat spacetime bicrossproduct model as the ambient noncommutative space. The derivation in \cite{Ma:bh} is outlined briefly in Section~3 to the point of the noncommutative wave equation (\ref{BHwave2}).

The present paper studies the black hole noncommutative wave operator in a certain scaling limit where distance $\rho$ in Planck lengths from the event horizon is  maintained while at the same time sending $\lambda_p\to 0$. This leads to a classical and much more amenable wave equation 
\[ \left( -2 \left( {\imath \over \nu}{\del\over\del t}+\rho \ln(1- {\imath  \over  \rho \nu }{\del\over\del t})\right)+{\del\over\del \rho}+\rho {\del^2\over\del \rho^2}\right)\psi=0\]
where the parameter $\nu$ is the Schwarzschild frequency. This still retains quantum gravity effects and has the same qualitative behaviour around the horizon as the original theory in \cite{Ma:bh}. Our findings are summarized at the end of the paper, see Figure~3. The physical significance  of  the novel `noncommutative skin' layer at the event horizon remains more fully to be understood. It is perfectly possible to read the paper starting at the scaling limit wave equation (\ref{BHwave3}).  

\section{Background to the bicrossproduct model quantum spacetime} 

We first explain some of the background to the model. Indeed, quantum groups arose out of physics in two different classes. The first and most well known are the Drinfeld-Jimbo quantum groups $U_q(\cg)$ associated to complex semisimple Lie algebras $\cg$ and coming out of integrable systems. Professor Jimbo at this conference rightly obtained the Weyl prize in part for that. The other less well-known class are the `bicrossproduct' quantum groups introduced by the author as associated to Lie group factorisations and coming out of ideas for quantum gravity and Planck scale physics\cite{Ma:pla}. At the time Planck scale physics was considered unreachable by experiment but in recent years this view has changed and these bicrossproduct quantum groups have gained in interest particularly as deformations of Poincar\'e and other inhomogenenous non-simple groups. 
 
 The general construction here has semidirect product form $C(M)\lrbicross U(\cg)$ where $M$ is a Lie group on which a Lie algebra $\cg$ acts and where $M$ acts back on $g$ and on the set of  its associated group $G$ so as to form a `matched pair'. The latter `back-reaction' induces a semdirect coproduct or, equivalently,  a dual quantum group $U(\cm)\bicross C(G)$, where $\cm$ is the Lie algebra of $M$, making everything symmetric between the two factors. There is also a canonical action of the first quantum group on $U(\cm)$ and its of dual on $U(\cg)$, which are therefore naturally arising noncommative geometries with these quantum groups as symmetries. Such data arise from the local factorisation of any group $X$ into $GM$ where neither subgroup need be normal, a very common occurrence in mathematics and physics.  
 
 As well as the general construction, including specifics for the Iwasawa factorisation, my 1988 PhD thesis explicitly covered\cite{Ma:pla,Ma:mat,Ma:hop} the case $SO(1,3)\approx SO(3)\CH_3$ where $\CH_3=\R^2\lcross\R$ is a curved (nonAbelian) version of $\R^3$. Its Lie algebra $h_3$ has relations (\ref{bicxt}) for  $i=1,2$, $t=x_3$ providing a noncommutative space on which the bicrossproduct quantum group $C(\CH_3)\lrbicross U(so_3)$ acts. The 4D case  studied a bit later in \cite{MaRue} is similarly based on $SO(2,3)\approx SO(1,3)\CH_{1,3}$ where the Lie algebra $h_{1,3}$ again has the relations (\ref{bicxt}) but with $i=1,2,3$. The associated bicrossproduct quantum Poincar\'e group $C(\CH_{1,3})\lrbicross U(so_{1,3})$ was shown to be isomorphic as an abstract Hopf algebra to a `$\kappa$-Poincar\'e quantum group' obtained by contraction of $U_q(so(2,3))$ in \cite{Luk} but had a different and inequivalent interpretation of the generators and therefore should not be confused with that. Moreover, the bicrossproduct model not only provided the correct interpretation of the Hopf algebra but  provided the quantum spacetime (\ref{bicxt}) on which it acts  as the enveloping algebra $U(h_{1,3})$. This is the bicrossproduct spacetime or `$\kappa$-Minkowski' model introduced by the author and H. Ruegg\cite{MaRue}.

Since then, the bicrossproduct  model quantum spacetime has attracted a  great deal of attention, not least due to its testable variable speed of light prediction\cite{AmeMa}. That there might be such a prediction had been hoped for on the basis that the deformed relations of the bicrossproduct Poincar\'e algebra involve a deformed form of the Casimir, but this of itself has no meaning as one can just rename the generators; merely calling a generator $P_0$ or $P_i$ does not make it energy or momentum. However, such speculations were turned into an actual prediction in \cite{AmeMa} by analysing the associated wave operator on noncommutative spacetime and its action on noncommutative plane waves $e^{\imath \vec k\cdot x}e^{\imath \omega t}$. We made the reasonable assumption that physical plane waves would, in this model of quantum gravity effects, be identified with these normal ordered mathematical plane waves. 
 
 Another feature of the bicrossproduct family is that the classical data of a matched pair of actions is entirely classical,  consisting in the Poincar\'e case of a certain non-linear action of the Lorentz group on the nonAbelian momentum group $\CH_{1,3}$ and vice versa. As a result the orbits of the Lorentz group in momentum space are classical but `squashed' into regions with limiting surfaces or accumulation boundaries. This phenomenon is identical to that first discovered in the $iso(3)$ case \cite{Ma:pla,Ma:mat,Ma:hop}, where the spherical orbits were deformed into non-concentrically nested spheres squashed into a region of $\R^3$ ; in the $iso(1,3)$ case the mass hyperbolae are similarly squashed into a cylinder $|\vec k|< \hbar/(c\lambda_p)$. In recent years some authors have sought to rebrand the bicrossproduct model as  `doubly special' because of this remarkable feature of squashed orbits and the feature of being able to refer everything to classical nonlinear equations, but without actually producing any genuinely new model with quantum group symmetry.  Both features are in fact typical of bicrossproduct models with noncompact factorising groups if one wants other examples, and I refer to \cite{Ma:qg2} for a recent review. Of perhaps more interest is how the bicrossproduct model emerges as a limit of a $q$-deformed geometry and how this in turn emerges from quantum gravity with cosmological constant, matters which are explored in the 2+1 case in \cite{MaSch}.

\section{Background to the noncommutative black hole wave operator}

 Given a manifold $M$, a metric $\bar g$ with Levi-Civita connection $\bar\nabla$ and a conformal Killing vector field $\tau$, we define in \cite{Ma:bh} 
a noncommutative spacetime coordinate algebra $C(M)\rtimes_\tau\R$ consisting of elements of the form $\psi=\sum\psi_n t^n$ where $\psi_n\in C(M)$ by which we mean, eg, smooth functions on $M$, and product defined via commutation relations
\[ [f, t]=\imath\lambda_p \tau(f),\quad\forall f\in C(M).\]
One can think of this as a cross product of $C(M)$ by the group $\R$ with action defined by $\tau$ and at least when $M$ is compact make everything precise as a cross product $C^*$-algebra. Next, given any pointwise invertible function $\beta\in C(M)$ we define a differential calculus as, by definition, a space of 1-forms $\Omega^1(C(M)\rtimes_\tau\R)$  spanned by usual 1-forms from $M$ along with two extra 1-forms $\extd t, \theta'$, with the commutation relations between functions and differentials\cite{Ma:bh}
\begin{eqnarray}\label{cr} [f,g]=0,\quad&& [f,t]=\imath\lambda_p \tau(f),\quad [\extd f,g]=\imath\lambda_p{\bar g}^{-1}(\bd f,\bd g)\theta'\nonumber\\
{} [\theta',f]=0,\quad [\theta',t]&=&\alpha\imath\lambda_p \theta',\quad  [\extd f,t]=\imath\lambda_p(\extd\tau(f)-\extd f)\nonumber\\  {} [f,\extd t]&=&\imath\lambda_p\extd f,\quad [\extd t,t]=\imath\lambda_p(\beta\theta'-\extd t).\end{eqnarray}
for all $g,f\in C(M)$, where $\alpha={2\over n}{\rm div}(\tau)-1$ if $M$ has dimension $n$. One of the Jacobi identities will quickly lead you to see that you need $\tau$ to be a conformal Killing vector. We do rather more in \cite{Ma:bh}, building $\Omega^1$ directly on the space of classical 1-forms $\bar\Omega^1(M)$ extended by two extra dimensions $\theta',\extd t$. The general principle of defining a differential structure via a bimodule of 1-forms (or an entire differential graded algebra) is common to most approaches to noncommutative geometry\cite{Con}.

We also consider 2-forms $\Omega^2(C(M)\rtimes\R)$ and show in \cite{Ma:bh} that the calculus is locally inner, i.e. in the neighbourhood of any point one can find a function $h\in C(M)$ such that
\[ \theta=\extd t- h\theta',\quad [\psi,\theta]=\imath\lambda_p\extd \psi,\quad \{\omega,\theta\}=-\imath\lambda_p\extd\omega\]
for all $\psi$ in the noncommutative coordinate algebra and all noncommutative 1-forms $\omega$.
Finally, we show that for all normal ordered $\psi\in C(M)\rtimes_\tau\R$ the exterior derivative has the form
\[ \extd \psi= \bd\psi + \del_0\psi\extd t+ \imath{\lambda_p\over 2}\theta'(\bDelta_{LB}-{1\over 2}\beta^{-1}\bar g^{-1}(\bd\beta))\psi+\imath\lambda_p\Delta_0\psi\theta'\]
where $\bd$ is the classical spatial exterior derivative, $\del_0$ is the finite difference operator
\[ \del_0 \psi(t)= {\psi(t)-\psi(t-\imath\lambda_p)\over\imath\lambda_p}\]
and $\bDelta_{LB}$ is the classical Laplace-Betrami operator on $M$. We use the inverse metric to view the classical 1-form $\bd\beta$ as a vector field, which we regard as a differential operator. Finally $2\Delta_0$ is a deformation of the Laplacian or double-derivative in the $t$ direction. We define the spacetime wave operator by
\[ \extd\psi=\bd\psi +\del_0\psi\extd t+\imath {\lambda_p\over 2}\square\psi\theta'\]
in keeping with a philosophy that the extra dimension $\theta'$ plays the role of the conjugate to the wave operator regarded as a `noncommutative vector field'. 

One can apply this to the spherically symmetric 3-geometry $(M,\bar g)$ underlying the Schwarzschild black hole. Here $\bar g=h(r)^2\bd r\btens\bd r+ r^2\bd^2\Omega$ where 
\[ h(r)^2= {1\over 1-{\gamma\over r}},\quad \gamma={2 G M\over c^2}\]
for a  black hole of mass $M$ centred at the origin. Then the classical black hole has the form of our static metric with 
\begin{equation}\label{betabh} \beta=-{1\over c^2 (1-{\gamma\over r})}.\end{equation}
Moreover, the 3-geometry has a conformal killing vector field $\tau={r\over h}{\del\over\del r}$. Hence by the above we have a quantum spacetime and a wave operator on it deforming that of a classical black hole.

As this wave operator and noncommutative geometry is a little hard to compute, we also propose\cite{Ma:bh} and in practice work with a more down to earth version of the theory where we let $M=\R^3$ with its conformal Killing vector field $\tau=r{\del\over\del r}$. Then the spacetime is the usual flat spacetime bicrossproduct model with relations (\ref{bicxt}). But for the calculus we take $\beta$ as in (\ref{betabh}) which is no longer the 5D flat space calculus\cite{Sit} for this model. Its wave operator is then not 
the flat space one used in \cite{AmeMa} but nor is it quite the black hole as we took a flat 3-geometry. To fix this we use the correct $\bDelta_{LB}-{1\over 2}\beta^{-1}\bar g^{-1}(\bd \beta)$ in place of the one computed from the flat metric. We call this the `minimally coupled black hole' as we basically adapted a flat space computation by replacing derivatives by covariant ones. 
Explicitly,
\begin{equation}\label{BHwave}\square_{BH}\psi(t)=2\Delta_0\psi(t)+\left(({2\over r}-{\gamma \over r^2}){\del\over\del r}+(1-{\gamma\over r}){\del^2\over\del r^2}+e_ie_i\right)\psi(t+\imath\lambda_p)\end{equation}
is our `minimally coupled' noncommutative black hole wave operator. Here $e_i={\del\over\del x_i}-{x_i\over r}{\del\over \del r}$ is the 2-sphere covariant derivative such that $e_ie_i=-{l(l+1)\over r^2}$ on the spherical harmonics $Y^l_m$. The nontrivial part is to compute $\Delta_0$. Let $L\psi(t)=\psi(t+\imath\lambda_p)$. Then\cite{Ma:bh}
\[ \Delta_0={ L\over c^2\lambda_p^2}\left( -X+{X^2\over 2}+(1-{\gamma\over r})(X-{\gamma\over r}\ln(1-{  X\over 1-{\gamma\over r}}))\right),\quad X=\imath\lambda_p\del_0.\]
If we denote the expression in large parentheses as $\CD(X)$ then the noncommutative wave equation becomes\cite{Ma:bh}
\begin{equation}\label{BHwave2}\left( {2\over c^2\lambda_p^2}{\mathcal D}(\imath\lambda_p\del_0)+({2\over r}-{\gamma \over r^2}){\del\over\del r}+(1-{\gamma\over r}){\del^2\over\del r^2}+e_ie_i\right)\psi=0\end{equation}
on normal-ordered $\psi$ with $t$-dependence to the right. We shall study this wave equation under the physical hypothesis that observed fields are to be identified in our model with normal ordered fields as a measure of quantum gravity corrections. 

Note that as $X\to 0$ we have $\CD(X)={X^2\over 2(1-{\gamma\over r})}+ O(X^3)$ so we obtain $-{1\over c^2(1-{\gamma\over r})}\del_0^2$ in this limit. This has the correct classical limit for the black hole wave equation. Also in the limit $r\to\infty$ we have $\CD(X)=-{X^2\over 2}+ O({\gamma\over r})$ so we obtain the standard bicrossproduct flat space wave operator used in \cite{AmeMa}, which replaces time derivatives by the finite difference ones $\del_0$.

However, the function $\CD(X)$ also has a nice limit $\CD(X)=-X+{X^2\over 2}+ O(r-\gamma)$ as $r\to \gamma$, i.e. close to the event horizon. So the noncommutative geometry as visible in the wave operator smooths out the coordinate singularity normally occurring there; in some sense it makes it one dimension better as a log in place of a $1/(r-\gamma)$. It similarly improves matters at the other boundary $r=\gamma e^{-\omega\lambda_p}$ which is the other side if an interregnum or `skin' region in which the logarithm has a negative argument. We will take a close look at both boundaries by means of a scaling limit. The geometrical picture that we refer to is shown in Figure~1 for positive and negative $\omega$.

\begin{figure}
\[ \includegraphics[scale=.5]{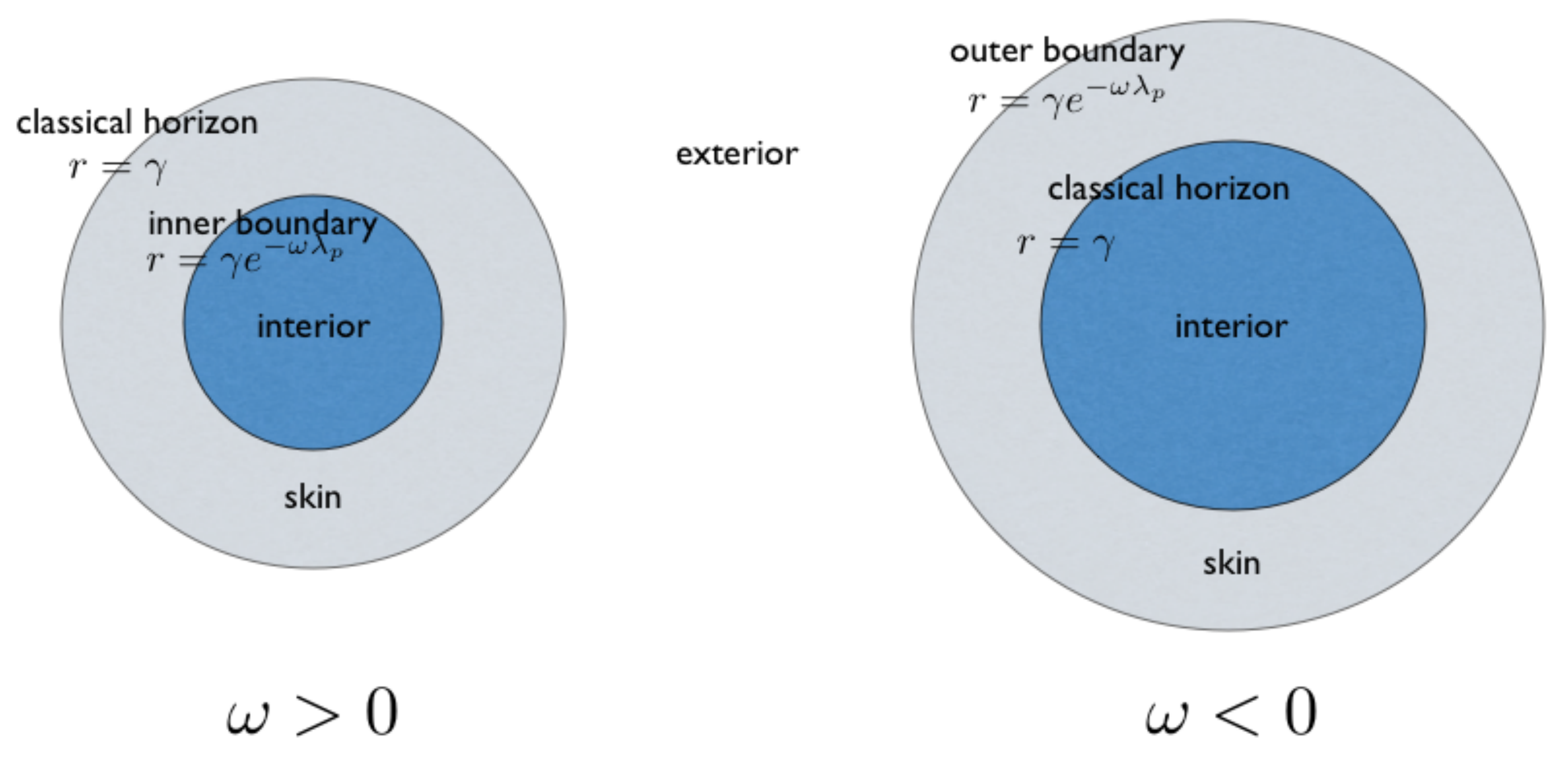}\]
\caption{Picture of the noncommutative black hole. The coordinate singularity at the event horizon is resolved into two boundaries with a frequency dependent  `interregnum' or skin between them of depth $\gamma|e^{-\omega\lambda_p}-1|$ where $\gamma$ is the Schwarzschild radius. The picture is different for positive and negative frequency $\omega$ modes.}
\end{figure}

\section{Solutions of the wave equation near the classical event horizon}

To study this limit $r\to\gamma$  further, we set
\[   r-\gamma= l_p\rho\]
where $l_p=c\lambda_p$ is the Planck length and $\rho$ is the distance from the horizon in Planck units. Then 
\[ {X\over 1-{\gamma\over r}}=\imath({\gamma+l_p\rho\over c\rho})\del_0.\]
Hence $r l_p$ times the wave equation (\ref{BHwave2}) becomes
\[\left( {2(\gamma +l_p\rho)\over l_p}{\mathcal D}(\imath({\gamma + l_p\rho\over c\rho})\del_0)+(1+{l_p\rho\over \gamma+l_p\rho}){\del\over\del \rho}+\rho{\del^2\over\del \rho^2}+ l_p(\gamma+\rho l_p) e_ie_i\right)\psi=0.\]
Although $l_p$ is what it is (about $10^{-33}cm$) we can simplify this equation from a mathematical point of view by setting $l_p\to 0$. This should be viewed as no more than a certain scaling limit approximation in which some quantum gravity effects are ignored but  a certain remnant remains. In this limit the wave equation in these coordinates becomes
 \begin{equation}\label{BHwave3}\left( -2 \left( {\imath \over \nu}{\del\over\del t}+\rho \ln(1- {\imath  \over  \rho \nu }{\del\over\del t})\right)+{\del\over\del \rho}+\rho {\del^2\over\del \rho^2}\right)\psi=0\end{equation}
if we assume bounded angular momentum so that  $l_p e_ie_i\to 0$. Here 
 \[ \nu= {c\over \gamma}\]
 is, roughly speaking, the frequency of a wave of wavelength the size of the black hole, as $\gamma$ is the Schwarzschild radius. 
 
 \begin{figure}\vskip -.5cm
 \[(a)\includegraphics[scale=.6]{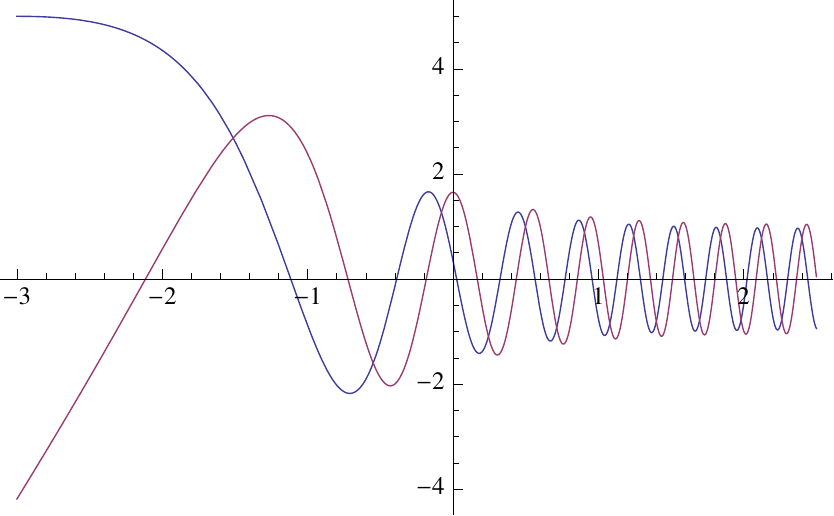}\]
 \[(b)\includegraphics[scale=.6]{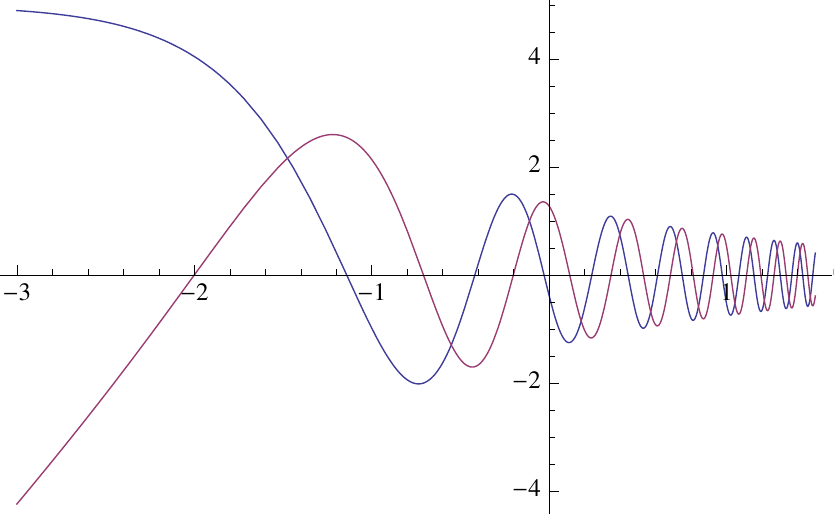}\includegraphics[scale=.6]{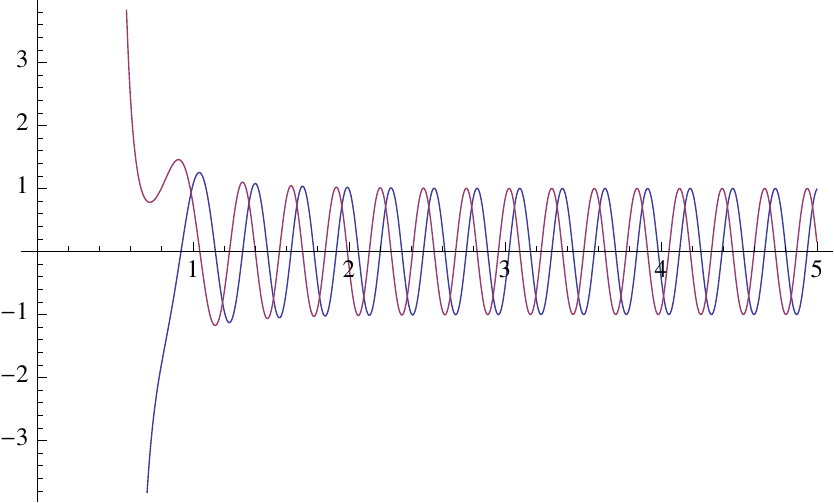}\]
 \[(c)\includegraphics[scale=.6]{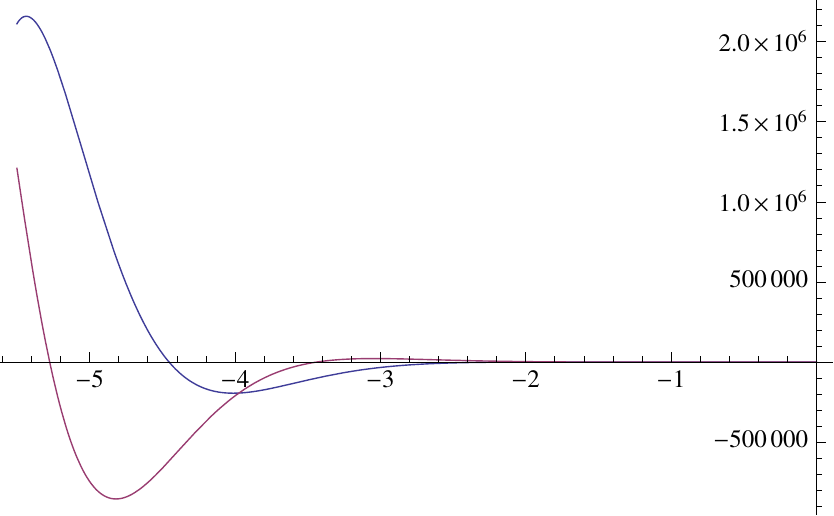}\includegraphics[scale=.6]{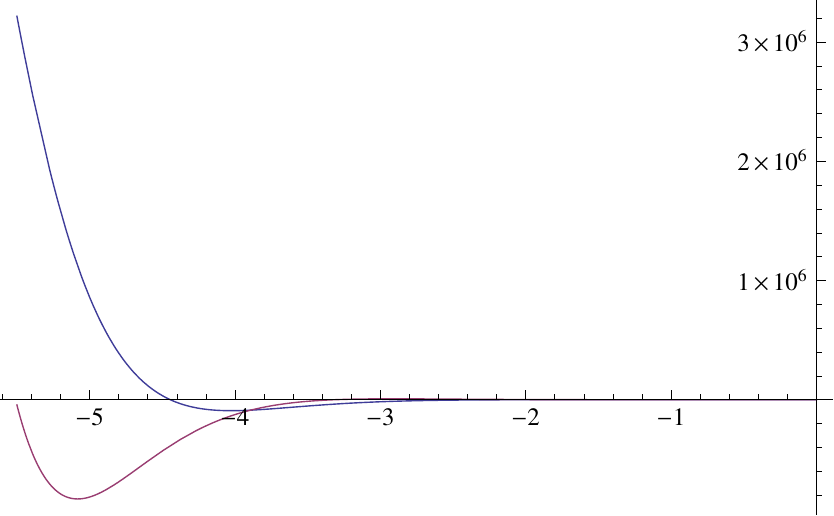}\]
 \[ (d)\includegraphics[scale=.6]{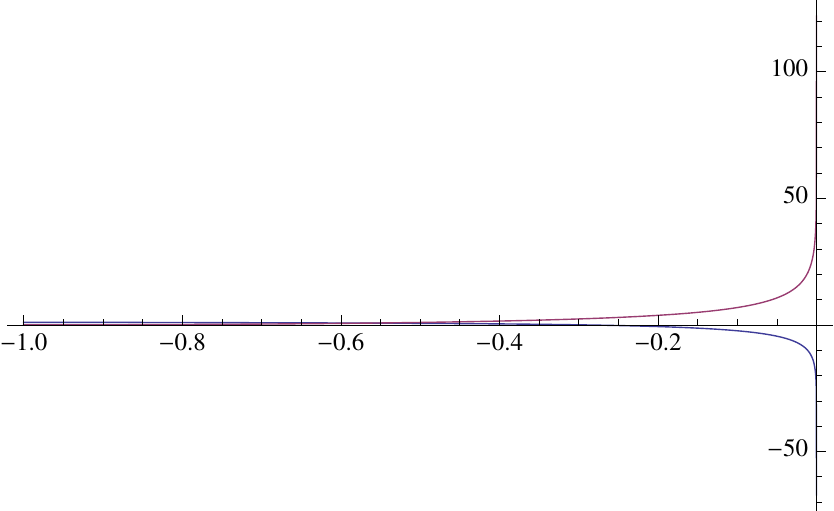}\]
 \[(e)\includegraphics[scale=.6]{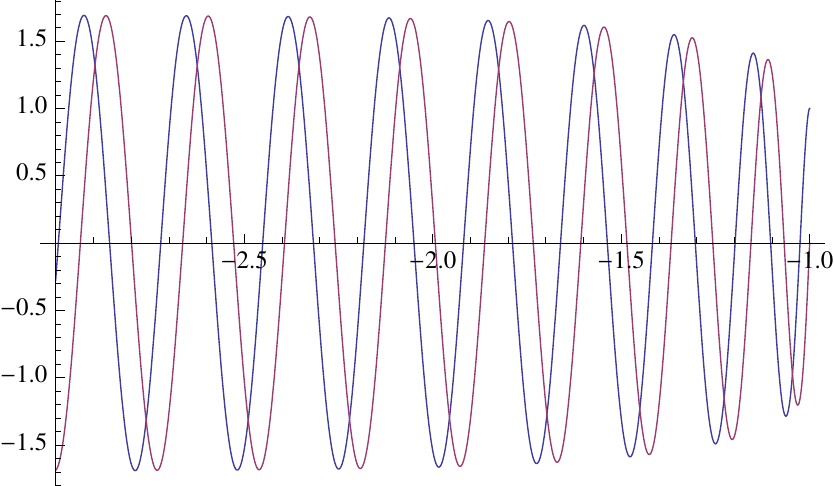}\includegraphics[scale=.6]{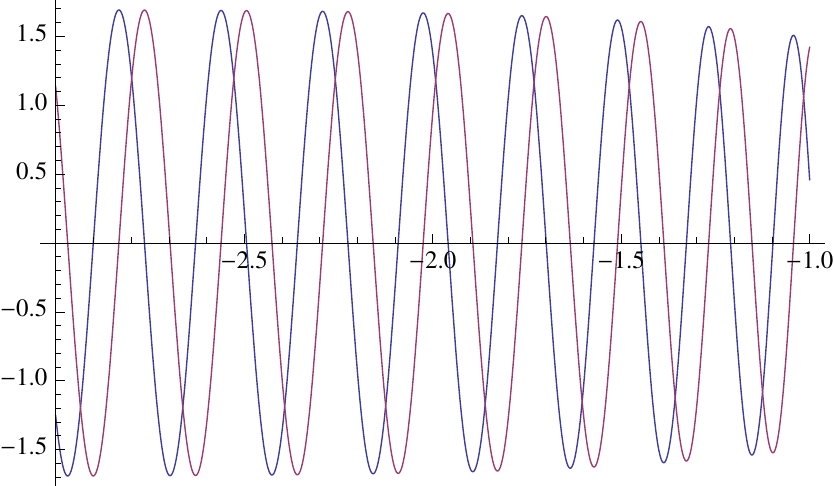}\]
 \caption{\small Scaling limit solutions against distance $\rho$ Planck lengths from event  horizon.  (a) Numerical  solution for ${\omega\over\nu}=10$ with differing boundary conditions from $\rho=10^5$  to $\rho=10^{-3}$  on log scale (b) Comparison with real and imaginary parts bessel solution for  $\rho>>{\omega\over\nu}=10^1$  (right)   and  $10^{-3}<\rho<<{\omega\over \nu}$  (left). (c) Real and imaginary parts  numerical solution (left) and comparison with hypergeometric solution (right) for  ${\omega\over\nu}=5.5$ in `skin' $-{\omega\over\nu}<\rho<0$  driven by real boundary condition at the horizon, on  linear scale. (d) Ditto numerical solution but for ${\omega\over\nu}=1$ driven by boundary condition at the interior boundary. (e) Numerical solution (left) for ${\omega\over\nu}=10$ with differing boundary conditions from $\rho=-10^1$ to $\rho=-10^3$, and comparison (right) with $|\rho|>> {\omega\over \nu}$ bessel solution, on log scale.}
 \end{figure}
We look at $\psi$ with time-dependence $e^{\imath \omega t}$ (separating variables). Then the equation (\ref{BHwave3}) becomes for each mode of frequency $\omega$,
 \begin{equation}\label{BHwave3w}\left( -2 \left( -{\omega \over \nu}+\rho \ln(1+{\omega \over  \rho \nu })\right)+{\del\over\del \rho}+\rho {\del^2\over\del \rho^2}\right)\psi=0\end{equation}
where $\psi$ is now a function just of $\rho$. Remembering the original physical interpretation of $\rho$, we see that for a given sign of $\omega$ there is a strip of width $|\omega|/\nu$ Planck lengths where the argument of the logarithm is negative. Specifically, this happens if   
\[ \omega>0:\quad  -{|\omega|\over\nu} < \rho<0,\quad \omega<0:\quad 0<\rho<{|\omega|\over \nu}.\]
Thus we have qualitatively the same kind of `quantum skin' or `interregnum' region in our scaling limit as for the original theory in Figure~1, but now with width $|\omega|/\nu$. This lies just inside the event horizon for positive frequencies and just outside it for negative frequencies. As long as we stay out of these regions, we have real solutions.  We use MATHEMATICA to compute numerical solutions, but our goal is to gain an analytic understanding of them in different regimes. We focus on $\omega>0$ but return to the negative frequency case at the end.

(i) {\em $\rho>>{\omega\over\nu}$ Far outside the event horizon} If we assume that
\begin{equation}\label{approx}{ |\omega| \over\nu} << \rho\end{equation}
then we can expand the logarithm. At lowest nontrivial order we obtain the conventional wave equation in unusual coordinates
\[ \left(-{1\over \rho \nu^2}{\del^2\over\del t^2}+{\del\over\del \rho}+\rho {\del^2\over\del \rho^2}\right)\psi=0.\]
Thus, if we write $\rho=e^{x\over\gamma}$ then the wave equation has solutions
\[ \psi_\pm(t,\rho)=e^{\imath \omega( t \pm {x\over c})}=e^{\imath\omega t}\rho^{\pm\imath{\omega\over\nu}}.\]
If we do a little better and expand to the next cubic order,  the wave equation takes the approximate form
\begin{equation}\label{BHcubic}\left(-{1\over \rho \nu^2}{\del^2\over\del t^2}-2{\imath\over 3 \rho^2 \nu^3}{\del^3\over\del t^3}+{\del\over\del \rho}+\rho {\del^2\over\del \rho^2}\right)\psi=0.\end{equation}
 This has solutions
 \[ \psi_\pm(t,\rho)=e^{\imath \omega t}  I_{\pm 2\imath{\omega\over \nu}}\left( \sqrt{8\omega^3\over 3\rho\nu^3}\right)\Gamma\left(1\pm 2\imath{\omega\over\nu}\right)\]
 where $I_n(z)$ is a bessel I function but extended to complex $n$ and the $\Gamma$ function is a normalisation. In another context one might consider taking $\imath\omega=\omega'$ real in which case note the poles in the normalisation factor of $\psi_\pm$ at $\mp$ positive integer $2\omega'\over\nu$.  Figure~2(b) on the right confirms, however, that the case of real $\omega$ is wave-like and indeed asymptotes to the plane waves just found when $\rho$ is large. Note that the plot is against log $\rho$ with ${\omega\over\nu}=10$.  The approximation (\ref{approx}) means that we can only approach the horizon to $\omega/\nu$  Planck lengths before quantum gravity effects render the weak field (i.e. small $\gamma$ or large $\nu$) expansion invalid. Thus, the right side of Figure~2(b) is only valid for $\rho>>10^1$ and this is confirmed on comparison with the numerical solution in Figure~2(a). 
 
 (ii){\em $0<\rho<<{\omega\over\nu}$ Just outside the event horizon}   The numerical solution in Figure~2(a) continues down to $\rho=10^{-3}$ or so on the log plot. At some point we enter the regime
 \begin{equation}\label{innerapprox} |\rho|<<{\omega\over\nu}\end{equation}
where the logarithm is suppressed and the wave equation becomes
\begin{equation}\label{BHnolog}2 {\imath  \over \nu}{\del\over\del t}\psi =\left({ \del\over\del \rho}+\rho {\del^2\over\del \rho^2}\right)\psi.\end{equation}
As a result the wave stops being so exponentially stretched in the direction of increasing $\rho$ (it begins to be stretched in the logarithmic plot) and starts to behave as a linear combination of 
\[ \psi\sim J_0\left(\sqrt{8 \rho\omega\over \nu}\right), \quad  Y_0\left(\sqrt{8 \rho\omega\over \nu}\right)\]
where $J_0,Y_0$ are bessel J and Y functions. These have respectively a finite value and zero slope at the event horizon $\rho=0$ or an infinite value and finite slope at $\rho=0$ in the logarithmic plot. Thus we conclude that the equation (\ref{BHwave3}) has a full set of solutions outside the event horizon becoming exponentially stretched plane waves for  large $\rho$ and tending to either a constant or a constant $\rho{\del\over\del \rho}\psi$ (the slope in the logarithmic plot) at the event horizon $\rho=0$. A generic boundary condition set at large $\rho$ will stimulate the latter $J_0'$ mode and result in a log divergence at the horizon, so this is presumably the fate of an infalling wave. However, a wave created at the event horizon (which would be relevant to Beckenstein-Hawking radiation) could take the form of the regular $J_0$ mode. 

(iii){\em $-{\omega\over\nu}<\rho<0$ In the skin layer} We now look  inside the event horizon. Initially, for small negative $\rho$ we are in the same regime as in (ii), except that there is now a term $-2\pi\imath\rho$ term in the wave operator if we take the standard choice of branch of the logarithm. This system can be solved using hypergeometric functions as 
\[ \psi\sim e^{\rho \sqrt{2 \pi\imath}} {}_1F_1\left(\frac{1}{2}+\frac{\omega}{\sqrt{2 \pi \imath}},1,-2 \rho \sqrt{2 \pi \imath}\right),  e^{\rho \sqrt{2 \pi\imath}} F_U\left(\frac{1}{2}+\frac{\omega}{\sqrt{2 \pi \imath}},1,-2 \rho \sqrt{2 \pi \imath}\right)\]
where $F_U$ is the confluent hypergeometric function.  We  focus for the moment on ${}_1F_1$ which has a finite value and zero slope   at the event horizon and can therefore match the $J_0$ (bessel J) on the other side. Its continuation further into the `skin' region acquires an imaginary part and both parts grow rapidly as we go deeper into the skin (more negative $\rho$).  The values at $\rho=-{\omega\over\nu}$ are (according to MATHEMATICA) given by a Laguerre L function and appear to grow very rapidly with $\omega\over\nu$. In Figure~2(c) we show this analytic approximate solution on the right and for comparison the full numerical solution, which is similar. The boundary condition is a real value 1 and zero slope at or close to $\rho=0$ and we have reduced to ${\omega\over\nu}=5.5$ in order to have a visible comparison at a point where significant `amplification' starts to appear.

For solutions driven by real conditions at (or rather, just inside) the interior boundary $\rho=-{\omega\over\nu}$ the numerical solution is shown in Figure~2(d) for a modest ${\omega\over\nu}=1$. In general a significant visible amplification in the numerical solution (say a factor of some $10^6$) sets in around ${\omega\over\nu}=4$ and a comparable imaginary part is similarly acquired, i.e. much as we had going the other way.  However, close inspection of the numerical solution appears to indicate that even for the example shown, there is in fact a log divergence at $\rho=0$. (This is more  visible in the middle scenario in our later Figure~3 where we have used a log scale so that a log divergence is a straight line and where we have suppressed a very large imaginary component in the plot allowing us to see a bit more detail than in Figure~2(d).) This brings us back to the hypergeometric mode $F_U$. This indeed has a log divergence in its real and imaginary part and a similar form in the region $-{\omega\over\nu}<<\rho<0$. Even though the approximation used breaks down as we approach $\rho=-{\omega\over\nu}$, as confirmed by the detail in Figure~3 mentioned, a generic boundary condition there can be expected to propagate into the region where the approximation (\ref{innerapprox}) is valid and thereafter drive this mode to produce a log divergence. The divergence can be matched in real slope at $\rho=0$ on the log plot to that of a $Y_0$ (bessel Y) mode outside, although the role of the similarly divergent imaginary part is not entirely clear. It is presumably related to the logarithmic asymptotic form. Overall, we see that there is a significant asymmetry in the modes crossing in the `skin layer' from the black hole interior, which appear to diverge as $F_U$ just inside the horizon, and $J_0$ modes crossing from outside the event horizon which are regular at the horizon and match to ${}_1F_1$ modes inside.

(iv){\em $\rho<<-{\omega\over\nu}$ Black hole interior} As we continue further with negative $\rho$ we return again to the regime $|\rho|>>{\omega\over\nu}$ and we are back to (\ref{BHcubic}), this time with $\rho<0$. If we change variables to $-\rho$ we have the same equation but with $-\omega$. Hence the cubic correction enters with the opposite sign. We have the same solutions Bessel  functions $I_n$ with imaginary $n$ as before but now an $\imath$ in the argument. As Figure~2(e) on the right confirms, we still have wave-like behaviour and these tend for large negative $\rho$ to plane waves in log plot. The difference is that whereas the bessel function in regime (i) showed a slight stretching of the waves in the log plot compared to large $\rho$, now we see a slight compression in the log plot compared to large negative $\rho$. We again see a good fit with the numerical solutions, shown for  real boundary conditions of finite amplitude/zero  derivative, or vice-versa. We have returned to ${\omega\over\nu}=10$ and have matched normalisation at large negative $\rho$.

(v){\em Negative frequencies} We have focussed on $\omega>0$. If we suppose instead that $\omega<0$ then note in (\ref{BHwave3w}) that we can change back to $\omega>0$ if we also change $\rho$ to $-\rho$. Hence the story for $\omega<0$ is exactly the same as above but with a left-right reflection in all of Figure~2. This time a mode entering from far outside the event horizon is slightly (but finitely) compressed compared to the overall frequency-independent compression expressed in the log plot, and diverges just above the classical event horizon. Meanwhile, a mode entering from the black hole `interior' is slightly (but finitely) stretched in the log plot, crosses the classical event horizon without coordinate singularity and is amplified as it passes out the other side. 

\begin{figure}
\[ \includegraphics[scale=.4]{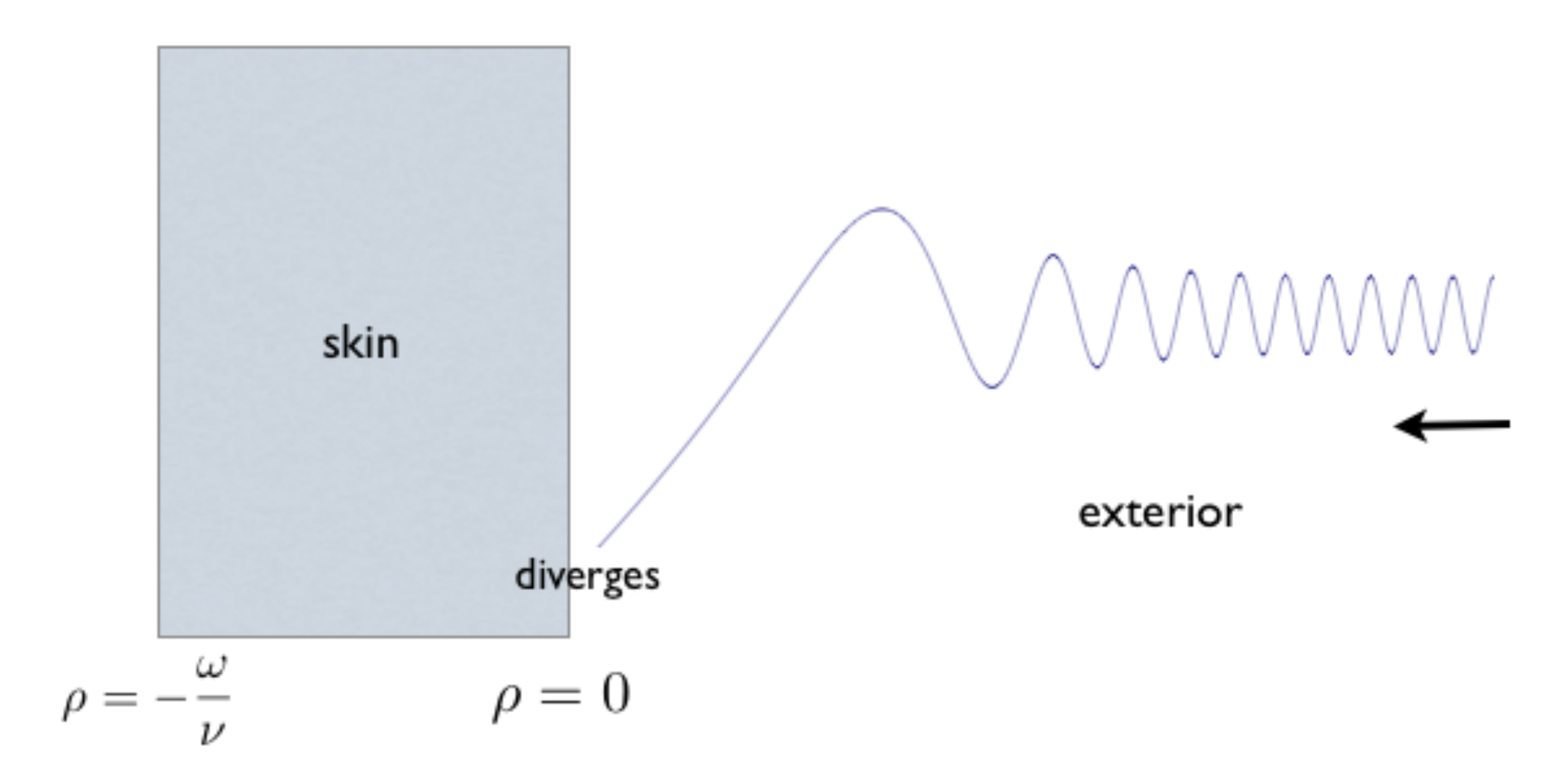}\]
\[ \includegraphics[scale=.4]{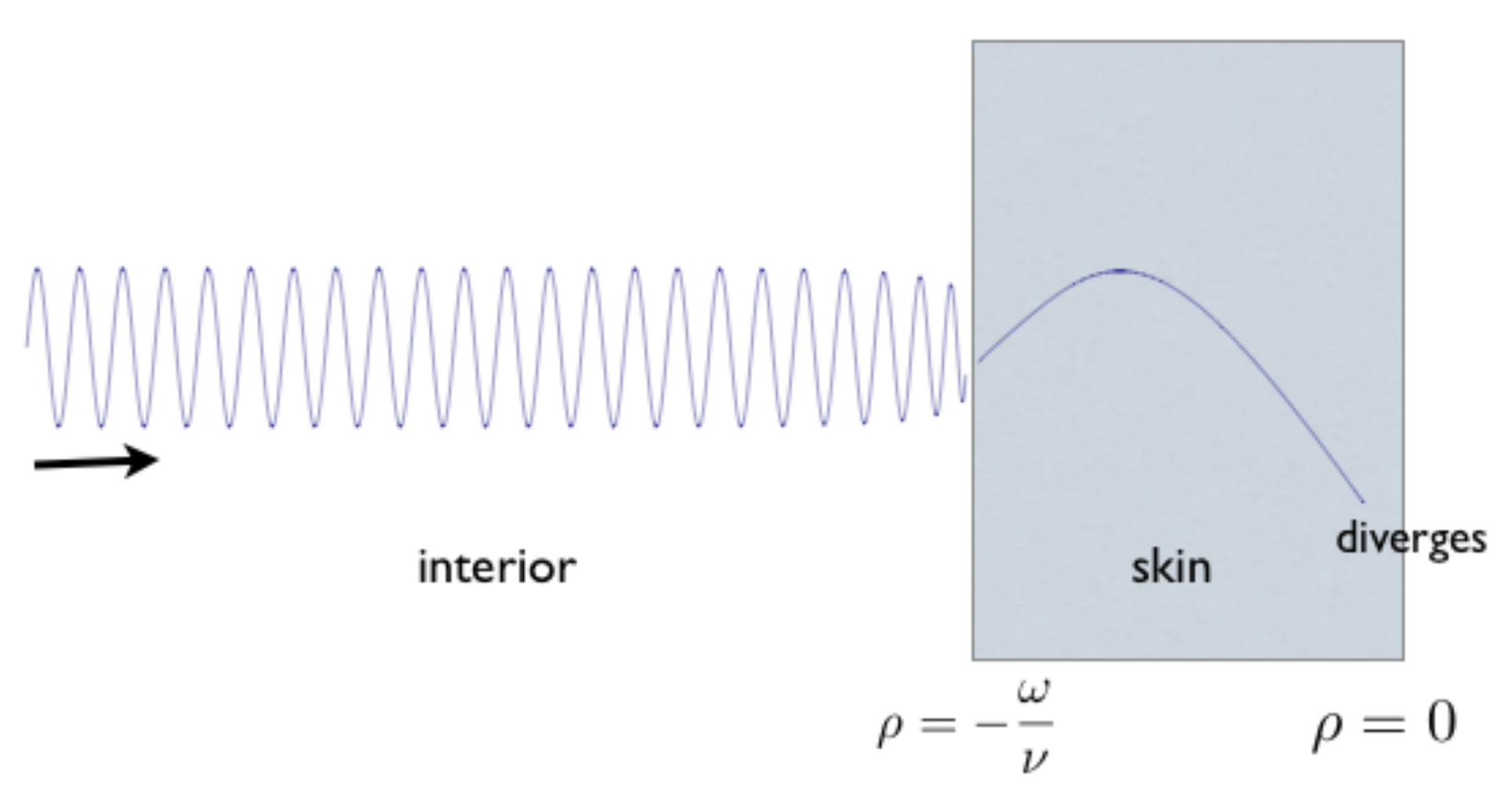}\]
\[ \includegraphics[scale=.4]{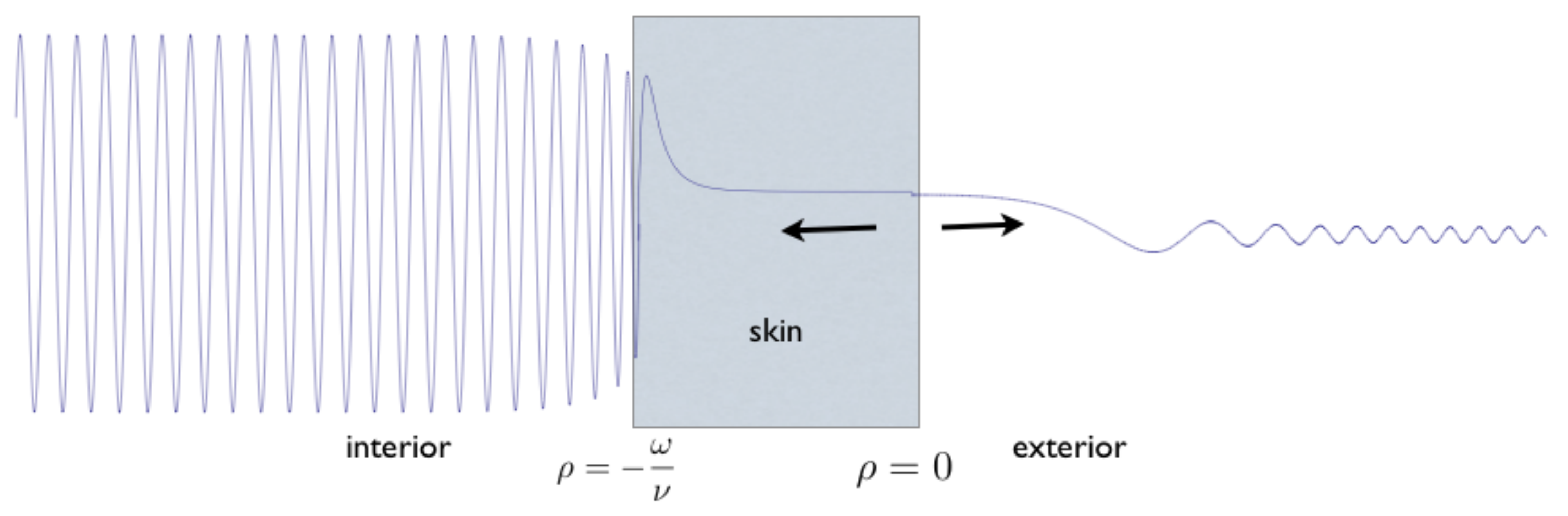}\]
\caption{Pastice of black hole wave solutions in scaling limit at the classical event horizon $\rho=0$ for  $\omega>0$ on a log plot. The system is driven by the location of the boundary conditions as indicated by the arrow. In the interior and exterior regions we have set ${\omega\over\nu}=10$ but have artificially reduced this in the skin regions to ${\omega\over\nu}=1.6$ in order to have a viable diagram with visible detail, and also note that solutions in the skin region acquire a comparable imaginary part which we omit for clarity. At top, we show a typical mode set far in the exterior, resulting in a log divergence but finite redshift just outside the horizon $\rho=0$. In middle we show a mode set in the `far' interior, resulting in a log divergence just inside the horizon.  At   bottom we show solutions set at the horizon. The pictures for $\omega<0$ are the same with `exterior' and `interior' interchanged, but note that the boundary condition illustrated would still be driving the skin region from the classical event horizon which is now just below the skin of the black hole.  }
\end{figure}

\bigskip 

All of this behaviour is remarkably similar to that of the full theory in \cite{Ma:bh}, suggesting that our simplified scaling limit wave  equation (\ref{BHwave3}) shares many qualitative features with the full wave equation. The only thing missing is that we cannot investigate `standing waves' in the black hole interior. This is because our scaling limit covers only the geometry near the horizon and a fixed number of Planck lengths from it (this includes the skin region). In effect, on entering the black hole interior the wave solutions in Figure~2(e) continue on towards the origin but this is considered as infinitely far away for a macroscopic black hole. On the other hand, one could consider a different scaling limit in which $\gamma=n l_p$ and $n$ is fixed as $l_p\to 0$ to analyse Planck size black holes in the model. This will be looked at elsewhere. 

Meanwhile,  the propagation in the two directions across the event horizon is summarised in Figure~3. Because the orbital angular momentum was washed out in our scaling limit and because we are far from the centre of the black hole, the system is more like propagation in a line. More precisely, two lines laid end to end. Thus, one could change variables to $x=\gamma\log \rho$ for $\rho>0$ and $y=-\gamma\log(-\rho)$ for $\rho<0$ and arrive at a more conventional $\del^2\over\del x^2$ or $\del^2\over\del y^2$ for the spatial part. We have refrained from this only in order to keep the original context. 
In the log plot in Figure~3 the interior and exterior propagations each span a factor of $10^6$ in $\rho$ and are set at ${\omega\over\nu}=10$. However, this would result in extremely high `amplification' factors in the skin region and hence for purposes of illustration only, we have set ${\omega\over\nu}=1.6$ for solution in the skin region. The boundary conditions are generally mixed ones setting amplitude and gradient, with the arrows indicating the location of the boundary condition and solving the equations from there, rather than indicating `direction of travel', which we have not particularly analysed. Although it does not appear to be too singular the equation at $\rho=-{\omega\over\nu}$ is numerically ill-conditioned and as a result the boundary conditions are set just inside the skin region rather than right on the boundary. 

We should also say that the physical interpretation of all of these features remains problematic. We have not discussed here issues such as unitarity (this is likely to be a problem) but it should be remembered that the entire noncommutative model is intended as an effective description of quantum gravity effects rather than a self-contained closed system. However, at least at first sight it would appear that waves from infinity are likely to log diverge at the classical event horizon $\rho=0$. We have seen that one can also create modes at the event horizon which are regular there and extend to waves outside. Inside the skin they grow rapidly and acquire an imaginary component, and the (now complex) solution appears to be matchable to waves continuing into the interior. This is shown in the lower part of Figure~3. Meanwhile, in the middle of Figure~3 we show waves in the interior which appear to be matchable to a solution inside which becomes complex and has a log divergence at $\rho=0$, this time on the inside of the classical event horizon. All of this is for positive frequency with a reversed picture for negative frequency. This represents a significant time-asymmetry, which is to be expected. Whether the log divergences at $\rho=0$ are important is not clear as they could be artefacts of the scaling limit; in the actual quantum gravity model presumably other physical effects would take over.

%\section*{References}


\begin{thebibliography}{99}
\bibitem{Ma:bh} Majid S 2010, Almost commutative Riemannian geometry, I: wave operators, arXiv:1009.2201 (math.QA), 50pp.

\bibitem{Ma:pla} Majid S 1988,  Hopf algebras for physics at the Planck scale, J. Class. Quant. Gravity 5, 1587-1607.

\bibitem{Ma:qreg} Majid S 1990, On q-regularization, Int. J. Modern Physics A. 5, 4689-4696


\bibitem{MaRue} Majid S\& Ruegg H 1994,  Bicrossproduct structure of the $\kappa$-Poincare group and non-commutative geometry, Phys. Lett. B. 334, 348-354

\bibitem{AmeMa} Amelino-Camelia G \& Majid S 2000, Waves on noncommutative spacetime and gamma-ray bursts, Int. J. Mod. Phys. A15, 4301-4323.


\bibitem{Ma:rie}Majid S 1999, Quantum and braided group Riemannian geometry, J. Geom. Phys. 30, 113-146


\bibitem{Ma:mat}Majid S 1990, Matched pairs of Lie groups associated to solutions of the Yang-Baxter equations, Pacific J. Math 141, 311-332.

\bibitem{Ma:hop}
Majid S 19991, Hopf-von Neumann algebra bicrossproducts, Kac algebra bicrossproducts, and classical Yang-Baxter equations, J. Functional Analysis 95, 291-319.

\bibitem{Luk}Lukierski J, Nowicki A, Ruegg H, \& Tolstoy VN 1991, $q$-Deformation of Poincar\'e algebra. Phys. Lett. B, 268,  331-338.


\bibitem{Ma:qg2} 
Majid S 2009, Algebraic approach to quantum gravity II: noncommutative spacetimes, in Approaches to Quantum Gravity, ed. D. Oriti. C.U.P. 466-492

\bibitem{MaSch}Majid S \& Schroers B 2009, q-Deformation and semidualisation in 3D quantum gravity, J. Phys
A 42, 425402 (40pp)

\bibitem{Sit} Sitarz A 1995, Noncommutative differential calculus on the $\kappa$-Minkowski space, Phys.Lett. B 349, 42-48.

\bibitem{Con}  Connes A 1994, {\em Noncommutative Geometry}, Academic Press.

\end{thebibliography}
\end{document}